\documentstyle[epsfig, twocolumn]{iso99}

\newcommand{\thepapername}{}

\begin{document}

\setlength{\parindent}{0pt}
\setlength{\parskip}{ 10pt plus 1pt minus 1pt}
\setlength{\hoffset}{-1.5truecm}
\setlength{\textwidth}{ 17.1truecm }
\setlength{\columnsep}{1truecm }
\setlength{\columnseprule}{0pt}
\setlength{\headheight}{12pt}
\setlength{\headsep}{20pt}
\pagestyle{veniceheadings}

\title{\bf ISO OBSERVATIONS OF SPIRALS: MODELLING THE FIR EMISSION}

\author{{\bf S.~Bianchi, P.B.~Alton, J.I.~Davies} \vspace{2mm} \\
Department of Physics \& Astronomy, Cardiff University, \\
PO Box 913, Cardiff CF2 3YB, U.K.
}

\maketitle

\begin{abstract}
ISO observations at 200$\mu$m have modified our view of the dust
component in spiral galaxies. For a sample of seven resolved spirals
we have retrieved a mean temperature of 20K, about 10K lower than
previous estimates based on IRAS data at shorter wavelengths.
Because of the steep dependence of far-infrared emission on the
dust temperature, the dust masses inferred from ISO fluxes are a
factor of 10 higher than  those derived from IRAS data only, leading
to gas-to-dust ratios close to the value observed in the Galaxy.
The scale-length of the 200$\mu$m emission is larger than for the IRAS
100$\mu$m emission, with colder dust at larger distances from the
galactic centre, as expected if the interstellar radiation field
is the main source of dust heating. The 200$\mu$m scale-length is also
larger than the optical, for all the galaxies in the sample.
This suggests that the dust distribution is more extended than that
of the stars.

A model of the dust heating is needed to derive the parameters of the
dust distribution from the Far-Infrared (FIR) emission. Therefore, we 
have adapted an existing radiative transfer code to deal with
dust emission. Simulated maps of the temperature distribution within the
dust disk and of the dust emission at any wavelength can be produced.
The stellar spectral energy distribution is derived from observations
in the ultraviolet, optical and near infrared. The parameters of
the dust distribution (scale-lengths and optical depth) are chosen
to reproduce the observed characteristics of the FIR emission, i.e. the
shape of the spectrum, the flux and the spatial distribution.
We describe the application of the model to one of the galaxies in the
sample, NGC~6946.
  \vspace {5pt} \\


  Key~words: ISOPHOT; spiral galaxies; extended emission; cold dust;
  FIR: observations; FIR: models.

\end{abstract}

\section{INTRODUCTION}

Diffuse dust is responsible for the internal extinction in a spiral galaxy.
The grains of this diffuse component are heated up to temperatures $T<20$\,K
by the mean galactic Interstellar Radiation Field (ISRF; \cite{sb_re95}). 
Emission from cold dust peaks at $\lambda> 150 \mu$m, while dust
in circumstellar regions is hotter and emits preferentially at shorter
wavelengths. The link between extinction and cold dust is revealed by
the similar values of the gas-to-dust mass ratio retrieved in the Galaxy
from independent studies of extinction and emission. Using mean dust
properties (\cite{sb_hi83};\cite{sb_wh92}), a value of 130 can be
derived from the correlation between the hydrogen column density and the 
B-V colour excess (\cite{sb_bo78}). From the Galactic FIR emission at
$\lambda>100$\,$\mu$m observed by the instrument DIRBE aboard the
satellite COBE, \cite*{sb_so94} derived a mean value of 160.

Until recently, the main source of FIR data for spiral galaxies has come
from the IRAS satellite. Because of its limited spectral coverage 
($\lambda<120\mu$m), IRAS was not able to detect cold dust. 
\cite*{sb_de90} measured the 
gas-to-dust ratio for a sample of 58 spiral galaxies, using the IRAS 
60\,$\mu$m and 100\,$\mu$m fluxes. A mean value of 1080 was derived, almost 
an order of magnitude larger than in the Galaxy. Assuming that the Galaxy 
is an average spiral, they explained the  discrepancy with 
90 per~cent of the dust mass being at $T\sim 15$\,K, too cold to be
detected by IRAS. 

Therefore, observations at longer wavelengths are essential to detect
cold dust and to trace the internal extinction in a spiral galaxy.
In this paper we describe ISOPHOT (\cite{sb_le96}) observations at
200\,$\mu$m of a sample of seven spiral galaxies. An extended cold
dust emission has been revealed (Section~\ref{sb_sec:iso}). 
A model of FIR emission has been built to derive the parameters of
the dust distribution from the observed properties of dust
emission (Section~\ref{sb_sec:mod}). A summary and a discussion
of the results are given in Section~\ref{sb_sec:fin}

\section{ISO OBSERVATIONS OF SPIRALS}
\label{sb_sec:iso}

We give in this section a summary of the ISOPHOT 200\,$\mu$m 
observations and results. A full description is presented 
in \cite*{sb_al98}. 

\begin{figure*}[!ht]
\begin{center}
\centerline{
\epsfig{file=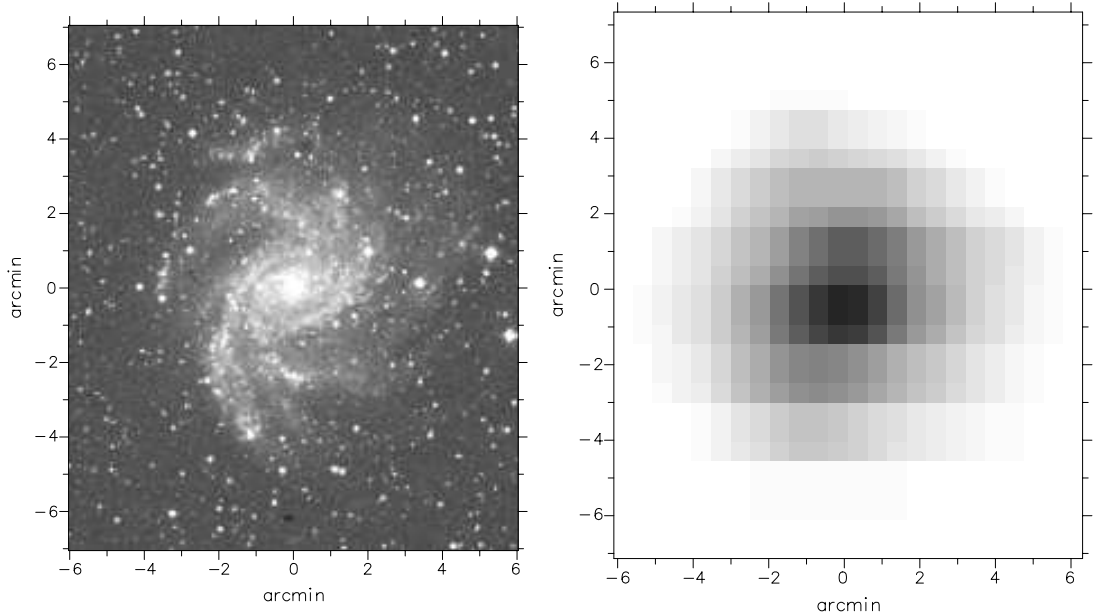,height=8.5cm}
}
\end{center}
\caption{\em Images of the spiral galaxy NGC6946, in the B-band (left) and 
at 200\,$\mu$m (right). Images are from Alton et~al. (1998)}
\label{sb_fig:ngc6946}
\end{figure*}

Eight objects were observed with ISOPHOT, seven spiral galaxies and one
irregular. The targets were selected to be representative of quiescent
objects with normal FIR-to-blue luminosity ratio ($L_{FIR}/L_B<1$,
\cite{sb_so87}) and to have large apparent sizes ($\sim$ 10 arcmin), to
be resolved by the ISO beam (FWHM=117 arcsec, \cite{sb_tu96}). B-band
images were also obtained, together with resolution enhanced HiRes
IRAS images at 60 and 100\,$\mu$m (\cite{sb_ri93}). The ISOPHOT images
are similar in morphology to the 100\,$\mu$m HiRes, that have a
comparable resolution. IRAS and B-band images were convolved to the ISO 
resolution and registered with the 200\,$\mu$m images.
As an example, we present in Figure~\ref{sb_fig:ngc6946} a
full-resolution B-band image and a 200\,$\mu$m image for one of the
objects of the sample, the spiral galaxy NGC~6946.

Total integrated fluxes at 60, 100 and 200\,$\mu$m were used to derive 
dust temperatures and masses. As for the dust emissivity, the form
$Q_{em}(\lambda)=Q_{em}(\lambda_0)\times(\lambda/\lambda_0)^{-\beta}$ 
was chosen, allowing for two different wavelength dependences, 
$\beta$=1 and 2, and
using $Q_{em}(\lambda_0)$ from  \cite*{sb_hi83}. After computing the
dust mass, gas-to-dust mass ratios were derived using the total 
(atomic+molecular) gas masses listed by \cite*{sb_de90}. Two set of
calculation were carried out: the {\em IRAS view}, using only the 60 and
100\,$\mu$m fluxes from IRAS; and the {\em ISO view}, using the 100\,$\mu$m 
IRAS flux and the  200\,$\mu$m ISO flux. The mean values of dust temperatures 
and gas-to-dust mass ratios over the observed sample are presented in 
Table~\ref{sb_tab:tempe}.

\begin{table}[!b]
\caption{\em Mean temperatures and gas-to-dust ratios for the observed
sample, derived for the {\em IRAS view} and the {\em ISO view}, for two
choices of the emissivity spectral index $\beta$. Values in brackets
allow for an overestimation of the ISO fluxes by 30 per~cent.}
\label{sb_tab:tempe}
\begin{center}
\leavevmode
\footnotesize
\begin{tabular}[h]{lrcc}
\hline \\[-5pt]
      &  $\beta$ & IRAS&   ISO\\[+5pt]
\hline \\[-5pt]
$\langle$T(K)$\rangle$       &  1&    34&    21 (24) \\
                 &  2&    28&    18 (19) \\
$\langle$Gas-to-Dust$\rangle$&  1&  2800&  230 (494) \\
                 &  2&  5100&  220 (446) \\
\hline \\
\end{tabular}
\end{center}
\end{table}

Clearly, a larger amount of cold dust is revealed using the 200\,$\mu$m
data.  Dust temperatures from the {\em ISO view} are colder by almost 10
degrees, and the gas-to-dust ratios are smaller by one order of magnitude,
with respect to the {\em IRAS view}. The values for the gas-to-dust ratio
are similar to the Galactic. The calibration of ISOPHOT is still uncertain.
An integrated flux at 200\,$\mu$m was derived for NGC~6946 from the 
observations of \cite*{sb_en91}. The ISO flux for the same galaxy is
larger by 30 per~cent. The same result is obtained comparing the observed
200\,$\mu$m background with that extrapolated from 100\,$\mu$m, using a 
Galactic spectrum at high latitude from \cite*{sb_re95}. However, even
accounting for an overestimation of the ISOPHOT calibration
(results in brackets in Table~\ref{sb_tab:tempe}), a larger quantity of
cold dust is revealed by the {\em ISO view}.

Uncertainties on the calibration do not affect the results about the 
spatial distribution of FIR emission. Azimuthally averaged profiles 
were obtained for the optical and FIR smoothed images, and the
exponential scale-lengths were measured between 1.5 arcmin and 3.5 arcmin.
The mean scale-lengths for the seven spirals in the sample are presented in
Table~\ref{sb_tab:scale}, normalised to the 200\,$\mu$m value.
The larger  200\,$\mu$m scale-length with respect to the other FIR
images shows that dust is colder at larger radii, consistently
with dust grains heated by a diffuse ISRF of higher intensity in the
centre of the galaxy.
The 60\,$\mu$m scale-length is similar to the 100\,$\mu$m, a result also 
observed in the Galaxy (\cite{sb_so89}) and explained with emission from
small grains whose heating conditions do not depend heavily on the ISRF
gradient.  The most striking results is the smaller B band scale-length
with respect to that at 200\,$\mu$m. If the dust distribution is similar
to the stellar, the scale-length of FIR emission should be smaller than 
the optical, because the dust temperature is higher in the centre of the 
disk. Therefore, the larger 200\,$\mu$m scale-length could be a hint for
a dust distribution more extended than the stellar. Extended dust
distributions also emerge from models of COBE emission at 140\,$\mu$m 
and at 240\,$\mu$m (\cite{sb_da97}) and from fits of surface brightness 
in edge-on galaxies (\cite{sb_xi99}).

\begin{table}[!t]
\caption{\em Mean scale-lengths of the sample}
\label{sb_tab:scale}
\begin{center}
\leavevmode
\footnotesize
\begin{tabular}[h]{cc}
\hline \\[-5pt]
60/200&  0.57$\pm$0.05 \\
100/200& 0.56$\pm$0.04 \\
B/200&   0.79$\pm$0.05 \\
\hline \\
\end{tabular}
\end{center}
\end{table}

To check for the influence of transient effects along the scan
direction, the galaxies were divided in four quadrants and the
scale-lengths analysis repeated only on the two quadrants
perpendicular to the scan direction. The same result was obtained.

\section{MODELLING THE FIR EMISSION}
\label{sb_sec:mod}

In this section we briefly describe a model of FIR emission in spiral
galaxies. The main purpose of the model is the derivation of the
geometrical and optical parameters of the dust distributions through
a comparison of simulations with observed FIR emission. In particular, we
wanted to test the hypothesis that the extended
dust emission described in Section~\ref{sb_sec:iso} is due to an
extended distribution of dust. 
A detailed description of the model and its results is deferred to a 
forthcoming paper (\cite{sb_bi99c}).

\subsection{Description of the model}
The model is based on a simplified version of the Monte Carlo radiative 
transfer code of \cite*{sb_bi96}. The Monte Carlo method allows for a correct 
treatment of multiple scattering in geometries typical of spiral galaxies.
For the models presented here, we use a double exponential disk geometry
for both stars and dust, although the code is able to work with other
distributions. The stellar and dust disk are geometrically defined by the
radial and vertical scale-lengths. The opacity of the dust disk is defined
by the face-on optical depth through the centre, $\tau_V$.
We used empirical dust properties for absorption and scattering
(\cite{sb_go97})

The code is monochromatic. Simulated images for a particular wavelength
are produced, together with a map of the energy absorbed by dust from
starlight at that wavelength.  Therefore, to obtain a map of the total energy 
absorbed by dust from the whole stellar Spectral Energy Distribution (SED),
simulations are produced for 17 wavelengths bands covering the spectral 
range of stellar emission, from the ionization limit to the Near-Infrared.

The main aim of this work is the modelling of the FIR emission, that is
produced by grains heated at thermal equilibrium. Since part of the absorbed 
energy is re-emitted in the Mid-Infrared by transiently heated small grains,
a correction is applied to subtract their contribution from the
absorbed energy map. The correction is derived from the absorption
efficiency of small grains in the dust model of \cite*{sb_dv90}.

From the corrected absorbed energy map the temperature distribution
along a meridian plane of the galaxy model is retrieved. A key quantity
for this calculation is the emissivity. We have derived the emissivity
value (\cite{sb_bi99}) from maps of Galactic FIR emission, temperature and 
extinction (\cite{sb_sc98}), assuming a spectral dependence as measured on
high signal-to-noise FIR spectra of the Galactic plane (\cite{sb_re95}).

Images of FIR emission are then produced from the temperature maps and
dust distribution, integrating along specific line of sights. In this
paper we describe an application to the spiral galaxy NGC~6946. Simulated
optical and FIR images have been produced for the inclination of 
34$^\circ$, appropriate for the galaxy (\cite{sb_at93}), for several
star-dust configurations. We have adopted a distance of 5.5 Mpc 
(\cite{sb_tu88}) for the galaxy; an intrinsic {\em unextinguished} stellar 
radial scale-length $\alpha_\star=2.5$\,kpc (derived from a K-band image);
a radial/vertical scale-lengths ratio $\alpha_\star/\beta_\star=14.4$
(a mean value derived from the compilation of vertical scale-lengths for 
different stellar types of \cite*{sb_wa92}).

\subsection{Results}

We first tried a model with a {\em standard} distribution for the dust disk,
i.e. with the dust radial scale-length of the same dimension as the
stellar ($\alpha_d=\alpha_\star$) and vertical scale-length one half of
the stellar ($\beta_d=0.5\beta_\star$). The choice of the dust vertical
scale-length is mainly dictated by the observation of extinction lanes in
edge-on galaxies, that cannot be explained with dust distribution
thicker than the stellar.

\begin{figure}[!t]
\begin{center}
\centerline{\epsfig{file=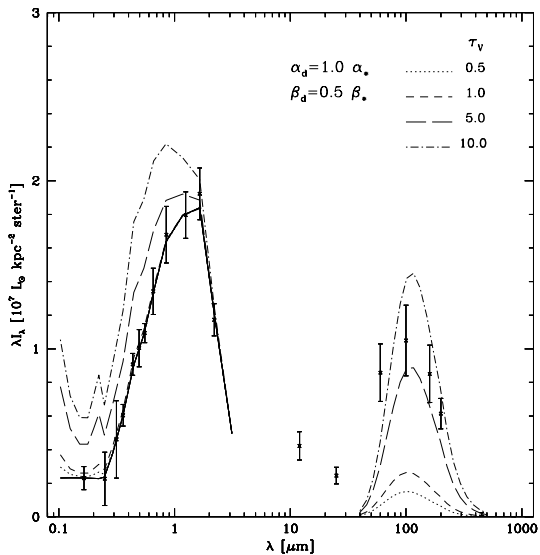,width=8.0cm}}
\end{center}
\caption{\em Optical and FIR SEDs for standard models 
($\alpha_d=\alpha_\star$, $\beta_d=0.5\beta_\star$) of different optical
depths.}
\label{sb_fig2}
\end{figure}

The SED of the stellar and dust emission
are presented in Figure~\ref{sb_fig2} for standard models of different
$\tau_V$. The optical data-points
for the stellar emission are from \cite*{sb_en91} and represent the flux
inside the B-band half-light aperture (5 arcmin diameter). UV fluxes inside
the same aperture are derived from \cite*{sb_ri95}. FIR fluxes inside
the same aperture are derived from IRAS HiRes images and from the 
200\,$\mu$m ISO image. The 160\,$\mu$m flux from \cite*{sb_en91} has
been included as well. The simulated optical and UV images were integrated
inside the B-band half light radius (also derived from the simulation)
and the fluxes normalised to the observed values. The solid line in
Figure~\ref{sb_fig2} represent this SED of the observed stellar
emission. The SED of the FIR emission (and that of the intrinsic,
unextinguished, stellar radiation) are presented for models of different
optical depths.

\begin{figure*}[!ht]
\begin{center}
\centerline{
\epsfig{file=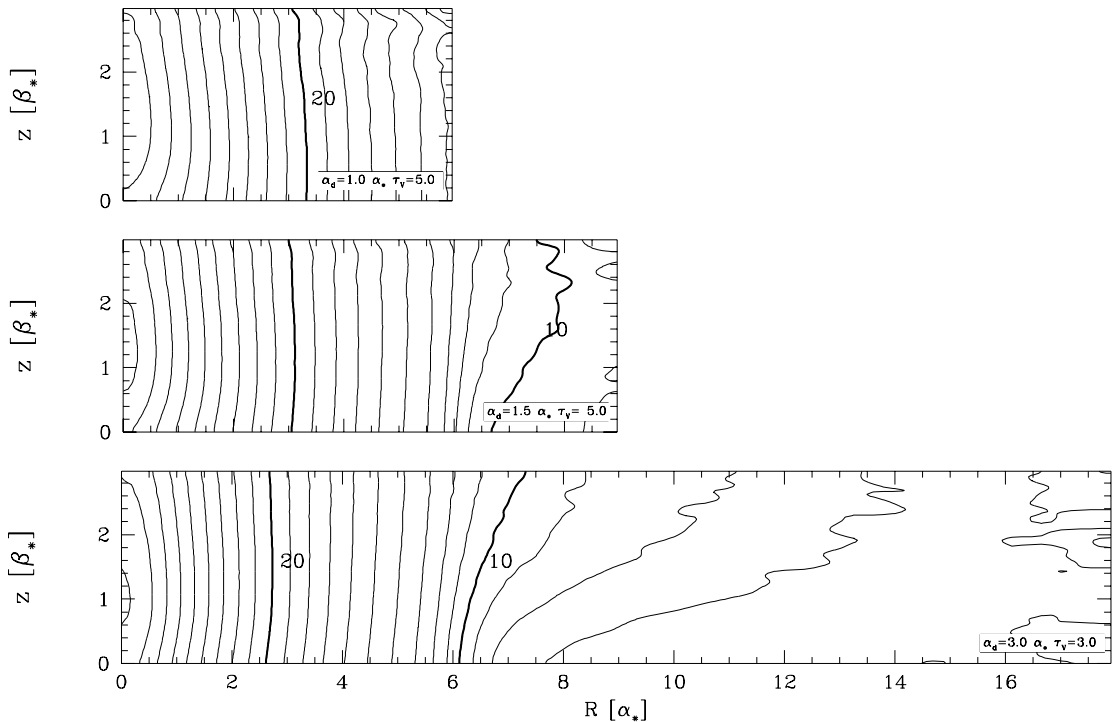,angle=90,height=10cm}
}
\end{center}
\caption{\em Temperature distribution on the meridian plane for the
standard model with $\tau_V=5$ (top panel) and the extended model with 
$\alpha_d=1.5\alpha_\star$ and $\tau_V=5$ (central panel). The bottom
panel presents the temperature distribution for the extended model with
$\alpha_d=3.0\alpha_\star$ and $\tau_V=3$. Contours are plotted every
1K and highlighted by a thicker line for the values of 10K and 20K. 
The scale on the x and y axes are different 
($\alpha_\star=14.4\beta_\star$). Dust and stellar disks are truncated
at six scale-lengths.}
\label{sb_fig5}
\end{figure*}

The SEDs of FIR radiation in Figure~\ref{sb_fig2} show that only
optically thick models ($\tau_V\sim 5$) can provide the observed amount
of energy absorbed (and emitted) by dust. As for the scale-length of FIR
emission, they increase with the optical depth. However, the B band
scale-length increases with extinction as well. Because of this, the
smaller value for the B/200 scale-lengths ratio occurs for optically thin
model with $\tau_V=0.5$ (1.4 vs 0.9 measured for NGC~6946). The
temperature distributions are quite similar for any of the dust disk
models. The temperature distribution on the meridian plane for the
standard model with $\tau_V=5.0$ is presented in the top panel of
Figure~\ref{sb_fig5}. The model temperatures are compatible with those
observed in the Galaxy (\cite{sb_so97}) and on NGC~6946 itself
(\cite{sb_da99}).

\begin{figure}[!t]
\begin{center}
\centerline{\epsfig{file=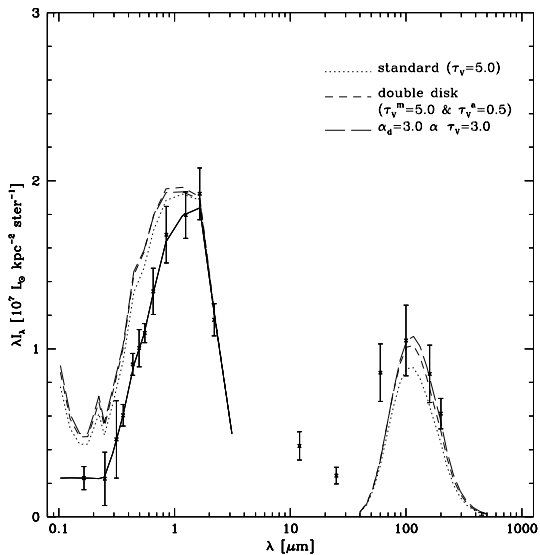,width=8.0cm}}
\end{center}
\caption{\em Optical and FIR SEDs for the double disk model, and for a
single disk model with $\alpha_d=3\alpha_\star$ and $\tau_V$=5.0.
The standard model with $\tau_V=5.0$ is also shown, for comparison.
}
\label{sb_fig4}
\end{figure}

\cite*{sb_xi99} modelled the surface brightness of a sample of seven
edge-on galaxies with a double exponential geometry for dust and stars.
The extinction lane is well fitted if the dust distribution is thinner
and has a larger radial scale-length than the stellar
($\langle\alpha_d/\alpha_\star\rangle=1.5, 
\langle\beta_d/\beta_\star\rangle=0.5$).  The dust
disk are found to be optically thin, with a mean $\tau_V=0.5$.
Figure~\ref{sb_fig4} shows the result for the case with extended disk.
Despite the internal extinction has increased, still optically thick
models are needed to absorb the right amount of energy. In an extended
model, larger amounts of colder dust are present in the outer skirts of
the disk (Figure~\ref{sb_fig5}) and this causes an increase in the
FIR scale-lengths. However, the increase in the B-band scale-length due 
to extinction compensates for the larger FIR scale-lengths and the B/200
scale-lengths ratios are larger then observed. The smaller value (1.15)
is again achieved for the optically thin case with $\tau_V=0.5$.
The models of COBE emission of \cite*{sb_da97} suggest that the dust
distribution is more extended then the stellar also in the vertical
direction ($\beta_d=2 \beta_\star$). The inclusion of a thicker dust
disk, however, does not modify the trend seen for the other models.

\begin{figure}[!t]
\begin{center}
\centerline{\epsfig{file=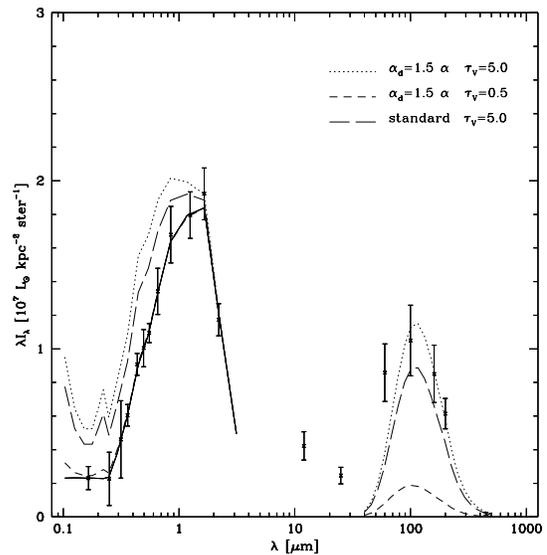,width=8.0cm}}
\end{center}
\caption{\em Optical and FIR SEDs for extended models 
($\alpha_d=1.5\alpha_\star$, $\beta_d=0.5\beta_\star$) of optical
depths $\tau_V=0.5$ and $\tau_V=5.0$. The standard model with
$\tau_V=5.0$ is also shown, for comparison.}
\label{sb_fig3}
\end{figure}

Since the molecular gas distribution of NGC~6946 follow closely the emission
detected by IRAS, \cite*{sb_da99} suggested that the broader 200\,$\mu$m
profile could be due to a more extended dust distribution associated
with the atomic gas. \cite*{sb_ta86} report the column density of
molecular and atomic hydrogen as a function of the galactocentric
distance. The molecular gas has a steep profile, with an exponential
scale-length $\approx$1\,$\alpha_\star$. The atomic gas distribution has
a dip in the centre, reaching a maximum at $\approx$2\,$\alpha_\star$,
then declining exponentially with a scale-length $\approx$3\,$\alpha_\star$.
The atomic and molecular phases have the same mass, but the former has a
much broader distribution. A model with two dust disk mimicking the gas
distributions is presented in Figure~\ref{sb_fig4}. As central face-on
optical depth we choose $\tau_V^m$=5 for the dust associated with the 
molecular gas, and $\tau_V^a$=5 for the dust associated with the atomic phase. 
The adopted optical depths were derived using the correlation between 
extinction and gas column density (\cite{sb_bo78}) and a standard extinction 
law for the Galaxy (\cite{sb_wh92}).

Despite the large extent of the disk associated with the atomic gas,
the FIR emission in the model is dominated by the optically thick disk 
associated with the molecular phase. The behaviour of the double disk model 
is therefore similar to that of a standard model with $\tau_V$=5. A good
fit can be found only for the SED while the B/200 scale-length ratio
is larger than observed. Interestingly, a better match of both FIR
emission and scale-lengths ratio can be found for a single disk model, of
the same scale-length of the one previously used for the atomic gas. 
An optical depth $\tau_V$=3 is needed to match the observed fluxes 
(Figure~\ref{sb_fig4}).  The bottom panel of Figure~\ref{sb_fig5}
shows the temperature distribution for this model: increasing the dust
scale-length with respect to the stellar results in larger amounts of
colder dust in regions of low intensity ISRF. The FIR scale-length has
increased enough to compete with the change in the B-band
scale-length due to extinction and the resulting B/200 scale-length ratio 
is now slightly smaller than 1. 

\section{SUMMARY \& CONCLUSIONS}
\label{sb_sec:fin}
ISO observations of a sample of seven spiral galaxies at 200\,$\mu$m have 
revealed large amounts of cold dust (\cite{sb_al98}). The FIR emission is 
found to have a larger scale-length than the optical 
($\langle\alpha_B/\alpha_{200}\rangle\approx 0.8$), thus suggesting a broader 
distribution for dust than for stars. 

We produced a model for the radiative transfer and FIR emission for 
geometries typical of a spiral galaxy. We applied
the model to the observations of NGC~6946. For all the model explored, an 
optically thick dust distribution (with central face-on optical depth 
$\tau_V\sim$5) is necessary to match the FIR spectrum. However,
optically thick models fail to reproduce the observed ratio of B-band
and 200\,$\mu$m scale-lengths. For models with dust radial scale-length 
larger than the stellar by a factor 1.5, as suggested by fitting of surface
brightness in edge-on galaxies, the optical-FIR scale-lengths ratio is
closer to the observed value only in optically thin cases. A better
fit of both FIR spectrum and spatial distribution of emission, in the
optically thick case, can be reached only for $\alpha_d/\alpha_\star\ge3$.

A few approximations have been adopted in this work. We have tested that
the results are non qualitatively modified by: the neglect of the
ionising stellar continuum; the presence of a small stellar bulge; 
the assumption of a single stellar vertical scale-length, instead of
having a thinner stellar disk for young objects (\cite{sb_bi99c})
emitting preferentially at smaller $\lambda$.
We have also assumed a single stellar radial scale-length at any
$\lambda$, following the
idea that colour gradients are mainly produced by dust extinction
(\cite{sb_pe95}).  \cite*{sb_dj96} claims that extinction cannot explain
the colour gradients observed in a sample of 86 spiral galaxy, for
{\em reasonable} models of galactic disks. However, this definitions
does not seem to include models with dust scale-lengths larger than the
stellar, that have colour gradients as large as the observed.

A major assumption in the radiative transfer model is that of a smooth
distribution, both for stars and dust. Clumping may be able to explain 
the paradox of needing both optically thick and optically thin distributions,
to reproduce the FIR spectrum and scale-lengths ratio, respectively.
Dust associated with the gas in molecular clouds may be
responsible for most of the FIR emission, if large amounts of energy are
absorbed from embedded stars. On the other hand, an optically thin
distribution correlated to the atomic gas will be able to explain the 
observed spatial distribution of FIR emission. The galaxy may look
as optically thin, when seen face-on, as derived from the surface 
brightness fits of \cite*{sb_xi99}.

Such a scenario, however, does not seem to be supported by observations.
For large objects with a well resolved clump structure, like the Milky
Way and M31, only a 30 per~cent of the FIR emission comes from star-forming
regions embedded in molecular clouds (\cite{sb_so97}; \cite{sb_xu96}).
It is then difficult to evaluate the effect of clumping on extinction and FIR 
emission. A study of the influence of clumping on the radiative transfer 
through a galactic disk is presented by \cite*{sb_bi99b}. It is found
that clumping affects only marginally the radiative transfer. It would
therefore be difficult for clumpy dust associated with molecular clouds
to pass ``undetected'' in surface brightness fits. However, results 
depend highly on the description adopted for the clumps distribution. A
description appropriate to the molecular distribution of NGC~6946,
rather than the Galactic adopted in \cite*{sb_bi99b}, may produce
different results. A proper model of FIR emission for a clumpy dust
distribution is therefore needed to confirm the prediction of the last
paragraph.

\end{document}